\documentclass[pra,onecolumn,floatfix,superscriptaddress,longbibliography,notitlepage, nofootinbib]{revtex4-1}
\pdfoutput=1

\usepackage[utf8]{inputenc}
\usepackage{braket}
\usepackage{amsmath}
\DeclareMathOperator{\Tr}{Tr}
\usepackage{dcolumn}   
\usepackage{bm}        
\usepackage{amssymb}
\usepackage{mathtools}
\usepackage{tikz}
\usepackage{qcircuit}
\usepackage{appendix}
\usepackage{hyperref}
\usepackage{subcaption}
\usepackage{amsmath,amsfonts,amssymb}
\usepackage{mathtools}
\usepackage{thmtools,thm-restate}
\usepackage{algorithm}
\usepackage{mathrsfs}

\usepackage{algpseudocode}

\newtheorem{definition}{Definition}

\newtheorem{lemma}{Lemma}

\newtheorem{method}{Algorithm}

\newcommand{\revise}[1]{\textcolor{black}{#1}}
\newcommand{\xbf}{\textbf{x}}

\newcommand{\Ibb}{\mathbb{I}}

\newcommand{\Rbb}{\mathbb{R}}

\usetikzlibrary{arrows.meta}

\def\BibTeX{{\rm B\kern-.05em{\sc i\kern-.025em b}\kern-.08em
    T\kern-.1667em\lower.7ex\hbox{E}\kern-.125emX}}

\begin{document}
\title{Simple Quantum Gradient Descent Without Coherent Oracle Access }
\author{Nhat A. Nghiem}
\affiliation{Department of Physics and Astronomy, State University of New York at Stony Brook, Stony Brook, NY 11794-3800, USA}
\affiliation{C. N. Yang Institute for Theoretical Physics, State University of New York at Stony Brook, Stony Brook, NY 11794-3840, USA}

\begin{abstract}
    The gradient descent method aims at finding local minima of a given multivariate function by moving along the direction of its gradient, and hence, the algorithm typically involves computing all partial derivatives of a given function, before updating the solution iteratively. In the work of Rebentrost et al. [New Journal of Physics, 21(7):073023, 2019], the authors translated the iterative optimization algorithm into a quantum setting, with some assumptions regarding certain structure of the given function, with oracle or black-box access to some matrix that specifies the structure. Here, we develop an alternative quantum framework for the gradient descent problem, based on the seminal quantum singular value transformation framework. We show that given only classical information of function of interest, it is possible to construct a quantum gradient descent algorithm with a running time logarithmical in the number of variables. In particular, our framework consumes exponentially less qubits than the prior quantum gradient descent algorithm and removes the need for any coherent oracle access to classical information. Thus, our work provides another example demonstrating the power of quantum singular value transformation framework, and in particular, it adds another instance revealing that quantum coherent access is not necessary for quantum computational advantage. 
\end{abstract}
\maketitle

\section{Introduction}
Quantum computing has made significant progress since its initial proposal \cite{feynman2018simulating, deutsch1985quantum, deutsch1992rapid, lloyd1996universal, shor1999polynomial, grover1996fast}. The fundamental principles of quantum mechanics, such as entanglement and superposition, enable a radically new paradigm for storing and processing information. One of the central questions in the field is understanding what quantum mechanical devices are capable of and how to harness their power effectively. The first aspect involves identifying computational problems that pose significant challenges for classical computers. The second focuses on developing systematic procedures—quantum algorithms—that quantum computers can execute. Similar to classical computing, quantum computation relies on specific resources, and an essential goal is to complete computational tasks as efficiently as possible, minimizing resource usage.

The study of quantum algorithms has gained considerable attention, leading to numerous breakthroughs in tackling computational challenges from different perspectives. Early milestones include Grover’s algorithm \cite{grover1996fast}, which demonstrated how a quantum computer could search an unstructured database, and Shor’s algorithm \cite{shor1999polynomial}, which showed that quantum computers could efficiently factorize numbers, raising concerns about the security of classical cryptographic systems. Additionally, \cite{feynman2018simulating, lloyd1996universal} introduced quantum simulation, which enables one quantum system to model the behavior of another. Since then, further developments, such as those in \cite{berry2007efficient, berry2012black, berry2014high, berry2015hamiltonian, low2017optimal, low2019hamiltonian, childs2022quantum}, have demonstrated that quantum computers can simulate a wide range of physical systems under specific assumptions. Building on quantum simulation techniques, Harrow, Hassidim, and Lloyd proposed the quantum linear system algorithm (HHL) \cite{harrow2009quantum}, later refined in \cite{childs2017quantum} using improved approximation methods. More recently, machine learning and data science have emerged as promising areas for quantum applications. Research by \cite{lloyd2013quantum, lloyd2014quantum, lloyd2016quantum, lloyd2020quantum, wiebe2012quantum, wiebe2014quantum} has explored how quantum computers can perform tasks such as classification, data fitting, and clustering—fundamental problems in modern machine learning.

The examples above highlight the potential applications of quantum computing. Each computational challenge requires distinct quantum techniques or algorithms tailored to its specific nature. A significant recent development in the field is the quantum singular value transformation (QSVT) framework \cite{gilyen2019quantum}, based on quantum signal processing \cite{low2017optimal, low2019hamiltonian}. This framework offers a unified perspective on quantum algorithms, revealing that many established algorithms—such as quantum walks \cite{childs2010relationship, ambainis2007quantum}, quantum search \cite{grover1996fast}, quantum linear solvers \cite{harrow2009quantum, childs2017lecture}, quantum support vector machines \cite{rebentrost2014quantum}, and quantum principal component analysis \cite{lloyd2013quantum}—can be understood using a common theoretical foundation. QSVT has not only simplified and unified quantum algorithm development but has also opened new avenues for quantum computational speedups. For instance, \cite{nghiem2022quantum} proposed a quantum algorithm for estimating the largest eigenvalue of a matrix given oracle access to its entries. Subsequent works \cite{nghiem2023improved1, nghiem2024improved} demonstrated that QSVT could significantly enhance eigenvalue estimation. Additionally, \cite{gilyen2022quantum} applied QSVT to implement the Petz recovery channel, a crucial tool in quantum information science.

\begin{figure}[h]
    \centering
    \begin{subfigure}[b]{0.4\textwidth}
        \centering
        \begin{tikzpicture}
    \draw[->] (-2, 0) -- (3, 0) node[right] {$x$};
    \draw[->] (0, -2) -- (0, 5) node[above] {$y$};

    \draw[domain=-1.5:2.0, smooth, variable=\x, blue, thick] plot ({\x}, {\x*\x}) node[right] {};

    \draw[domain=-0.5:2.0, smooth, variable=\x, red, thick] plot ({\x}, {2*\x - 1}) node[below] {};

    \filldraw[black] (1, 1) circle (2pt) node[above right] {$(1, 1)$};

    \node at (2, 4) [blue] {$y = x^2$};
    \node at (2, 3) [red] {Tangent line};

\end{tikzpicture}
    \caption{Function $y=x^2$ }
    \label{fig: fig1}
    \end{subfigure}
    \hfill
    \begin{subfigure}[b]{0.4\textwidth}
    \centering
    \begin{tikzpicture}
    \draw[->] (-2, 0) -- (3, 0) node[right] {$x$};
    \draw[->] (0, -2) -- (0, 5) node[above] {$y$};

    \draw[domain=-3.2:3.2, smooth, variable=\x, blue, thick] 
        plot ({\x}, {sin(2*\x r) + 0.3*\x*\x}) 
        node[above] {$y = \sin(2x) + 0.3x^2$};

    \foreach \x in {-2.35, -1.57, 0, 1.57, 2.35}{
        \filldraw[black] (\x, {sin(2*\x r) + 0.3*\x*\x}) circle (1.5pt);
    }


\end{tikzpicture}
\label{fig: fig2}
\caption{Function $y = \sin(2x) + 0.3x^2$}
    \end{subfigure}
\caption{ Simple illustration for the gradient descent method. It begins at some initially random points $\xbf_0$, then gradually ``slides'' along its gradient descent and move toward the local minima. In the left figure above, as the function $y=x^2$ is convex, the local minima is also global minima, and gradient descent is guaranteed to converge to such minima. On the right figure, the function is more complicated, exhibiting many local minima. Hence, the result of gradient descent depends on the initial guess $\xbf_0$, i.e., different initial guess leads to different minima. }
\label{fig: mainfigure}
\end{figure}
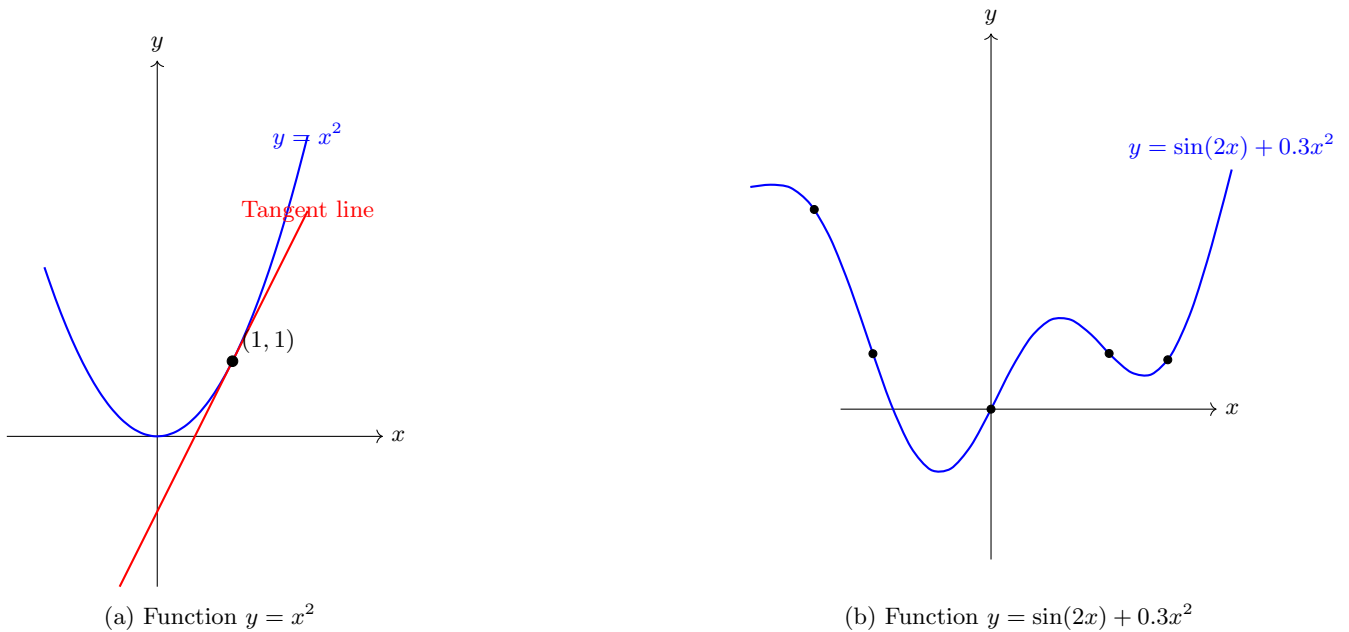

Encouraged by these successes, researchers have sought to expand the scope of QSVT. Surprisingly, QSVT exhibits greater simplicity and power than initially anticipated. A key insight from \cite{gilyen2019quantum} is that QSVT excels at manipulating polynomial functions, particularly those involving block-encoded operators. This observation has motivated investigations into problems related to polynomials, including quantum approaches to gradient descent \cite{rebentrost2019quantum}. Gradient descent is a fundamental optimization technique used to locate local minima of multivariate functions. Its principle is straightforward: at a local minimum, the gradient (i.e., the derivative) of the function is zero. The method iteratively adjusts an initial guess $\xbf_0$ by following the steepest descent direction until convergence, where further iterations produce negligible changes (see Figure \ref{fig: mainfigure} for an illustration in a simple setting).
The aforementioned example can be generalized to multivariable setting in a very straightforward manner. The core computational routine of gradient descent method is the evaluation of gradient of given function. Hence, it requires the ability to compute the partial derivative with respect to all variables. In principle, given the ability to compute, or evaluate a function at a given point and its surrounding, then all partial derivative could be numerically approximated by the simple derivative formula. In \cite{rebentrost2019quantum}, the authors considered the case where the function of interest is a special kind of polynomial, so-called homogeneous polynomial of even degree. Such polynomial admits a specific algebraic structure, which allows the gradient to be explicitly written down, without numerical approximation (to be elaborated later). Given an oracle, or blackbox access to entries of a certain matrix, the Ref.~\cite{rebentrost2019quantum} showed that quantum computer could perform gradient descent polylogarithmically in the dimension of solution vector, which could be very beneficial for high-dimensional problem. 

In this work, we draw on some insights, such as the suitability of quantum singular value transformation with respect to polynomial functions. Additionally, we are inspired by the proposal of \cite{rebentrost2019quantum}, that not only a homogeneous polynomial of even degree, but any monomial admits a simple analytical form for the partial derivative and hence the gradient. Furthermore, it is also straightforward to see that any more complicated polynomial function could be constructed from monomials, and the chain rule allows the gradient of such function to be written as the sum of the gradient of individual monomials. Thus, it is sufficient to operate with monomials, for which QSVT turns out to be a very good tool. We shall show that, by incorporating simple techniques from QSVT \cite{gilyen2019quantum}, we can construct a quantum algorithm for gradient descent, which can be performed with a broader class of polynomials compared to \cite{rebentrost2019quantum}. In particular, our method does not assume the coherent oracle or black-box access, which makes the method much more approachable for near-term realization. 

The structure of this work is as follows. In section \ref{sec: review}, we provide a formal overview of gradient descent problem and briefly describing the Ref.~\cite{rebentrost2019quantum}. Next, in section \ref{sec: newmethod}, we outline two quantum gradient descent algorithms, each suitable to different types of functions. In the following Section \ref{sec: discussion}, we discuss our framework from a broader context, in particular, we show how our work performs relative to that of the original proposal \cite{rebentrost2019quantum} and best-known classical method, emphasizing the advantage of our proposal. Conclusion as well as brief suggestion for future exploration is provided in Section \ref{sec: conclusion}.

\section{Review of Gradient Descent and Previous Work}
\label{sec: review}
Gradient descent is a very popular method designed for optimization problems. In such a problem, we are given an objective function $f: \Rbb^n \rightarrow \Rbb$ of $n$ variables, and the goal is to find a point $\xbf \equiv (x_1,x_2,...,x_n)$ where the value of the given function is minimized. The algorithmic procedure of gradient descent is quite simple. We begin with some random guess $\xbf_0$, then iterate the following procedure:
\begin{align}
    \xbf_{t+1} = \xbf_t - \eta \bigtriangledown f(\xbf_t) 
    \label{eqn: 1}
\end{align}
where $\eta >0$ is a hyperparameter. The iteration procedure typically continues until $\xbf$ reaches a ``fixed'' point, where more iterations no longer produce a significant change, indicating that the gradient $\bigtriangledown f(\xbf_t)$ approaches 0, and hence the function $f$ approaches its minimum value. Classically, it is apparent that the complexity of the gradient descent algorithm depends on how the function $f$, as well as its gradient, is evaluated, plus the total number of iterations. In the work of Rebentrost et al. \cite{rebentrost2019quantum}, the author considered the gradient descent problem in a quantum setting and provided a quantum algorithm to perform gradient descent. More specifically, they consider $f$ to be a homonogeous polynomial of even degree $2p$, which admits the following tensor algebraic form:
\begin{align}
    f(\xbf) = \frac{1}{2} (\xbf^T)^{\otimes p} A (\xbf)^{\otimes p}
    \label{eqn: tensorformf}
\end{align}
where $A$ is a matrix of size $n^p \times n^p $ of bounded matrix norm $|A| = \Gamma$. Furthermore, they also assume that $A$ is sparse (with sparsity $s$) and that its entries can be accessed through an oracle. Furthermore, they imposed a spherical constraint, where, at any iteration step, the solution is normalized: $\xbf \longrightarrow \ket{\xbf} \equiv \xbf/|\xbf|$. The above tensor structure of the objective function $f$ allows the gradient to be be conveniently written as:
\begin{align}
    \bigtriangledown f(\xbf) = D(\xbf) \xbf
\end{align}
where $D(\xbf)$ is an $n \times n$ matrix, and in principle $D$ could be obtained from $A$ (see section 2.1 in \cite{rebentrost2019quantum} for more details). While the algorithm of \cite{rebentrost2019quantum} is far more detailed and complicated, we capitulate their main idea and flow here: oracle access to $A$ allows them to use the tool from quantum simulation \cite{berry2007efficient, berry2012black, berry2015hamiltonian} to simulate $\exp(-iAT)$ for some $T>0$, then they use such exponentiation $\exp(-iA T)$ the method introduced \cite{lloyd2013quantum} to construct the operator $\exp(-iD(\xbf_t)T)$, and finally, they use method of \cite{wiebe2012quantum} to construct the operator $D(\xbf_t)$, or more precisely, they are able to obtain $D(\xbf_t) \xbf_t$, which is then used to update the temporal solution $\xbf_t$ according to equation \ref{eqn: 1}. The spherical constraint could be simply executed by performing a measurement (on some other ancillary system) and they would obtain a state, or normalized version of $\xbf_t - \eta \bigtriangledown f(\xbf_t)$, which is exactly $\ket{\xbf_{t+1}}$. Now we proceed to describe our new approach to (quantum) gradient descent. We also lift the spherical constraint, and treat the problem in its full generality. 
 
\section{Alternative Framework for Quantum Gradient Descent}
\label{sec: newmethod}
\subsection{Generic case}
\label{sec: genericfunction}
Suppose that the objective function $f: \Rbb^n \rightarrow \Rbb$ of our interest is:
\begin{align}
    f(x_1,x_2,...,x_n) = \sum_{i=1}^K a_i f_i(x_1,x_2, ..., x_n) 
\end{align}
where each $f_i(x_1,x_2, ..., x_n) $ is a monomial and $a_i$ is some real coefficient. Without loss of generality, we assume further that our domain of interest is $\mathscr{D} = [-1/2,1/2]^n$ and $|f(x_1,x_2,...,x_n)| \leq 1/2$ in this domain $\mathscr{D}$. Furthermore, we assume that the norm of gradient of $f$ is bounded by some known value $M$, i.e., $| \bigtriangledown f(x_1,x_2,...,x_n) | \leq M $ for all $(x_1,x_2,...,x_n) \in \mathscr{D}$.

For each  $f_i(x_1,x_2, ..., x_n) $, let $v(f_i)$ denotes the number of variables of $f_i$, $s(f_i)$ denotes the set of such non-zero variables, and $d(f_i)$ denotes the total degree of $f_i$. For example, if $f_i(x_1,x_2,...,x_n) = x_1^2 x_2^3 x_5$ then $v(f_i) = 3$, $s(f_i) = (x_1,x_2,x_5 )$ and $d(f_i) = 6$. Likewise, if  $f_i(x_1,x_2, ..., x_n) = x_1 x_2 x_3 x_7^4 x_9^2$ then $v(f_i) = 5$, $s(f_i) = (x_1,x_2,x_3,x_7,x_9 ) $ and $d(f_i) = 9$. In the following, we set a convention that:
\begin{align}
    f_i(x_1,x_2, ..., x_n) = x_1^{i_1} x_2^{i_2} ... x_n^{i_n}
\end{align}
where $i_1,i_2,...,i_n \in \mathbb{Z}$. Then it is simple to see that $| s(f_i)| = v(f_i)$, $\max v(f_i) = n$ and that if $v(f_i) < n $, then there are $n-v(f_i)$ values of those among $i_1,i_2,...,i_n$ must be zero. 

We first mention an essential tool for our subsequent construction:
\begin{lemma}[Theorem 2 in \cite{rattew2023non}; \revise{Block Encoding into a Diagonal Matrix}]
\label{lemma: diagonal}
     Given an $\log(n)$-qubit quantum state specified by a state-preparation-unitary $U$, such that $\ket{\psi}_n=U\ket{0}_n=\sum^{n}_{k=1}\psi_k \ket{k-1}_n$ (with $\psi_k \in \mathbb{C}$), we can prepare an exact block-encoding $U_A$ of the diagonal matrix $A = {\rm diag}(\psi_1, ...,\psi_{n})$ with $\mathcal{O}(\log(n))$ circuit depth and a total of $\mathcal{O}(1)$ queries to a controlled-$U$ gate  with $\log(n)+3$ ancillary qubits.
\end{lemma}
Given some unitary $U$ that generates the state $\sum_{j=1}^n x_j \ket{j-1}\bra{j-1}$, it allows us to construct the (exact) block encoding of the diagonal operator:
\begin{align}
    \textbf{X} = \begin{pmatrix}
        x_{1} & \cdot & \cdot & \cdot \\
        \cdot & x_{2} & \cdot & \cdot \\
        \cdot & \cdot & \ddots & \cdot \\
        \cdot & \cdot & \cdot & x_{n}
    \end{pmatrix}
\end{align}
Let $\ket{j}$ $(j=0,1,...,n-1)$ denote the arbitrary computational basis state of a Hilbert space of $\log(n)$ qubits. It is easy to see that one just needs $X$ gates to construct arbitrary $\ket{j}$ from $\ket{0}$. For example, to construct $\ket{0}^{\otimes (\log(n)-2)} \otimes \ket{11}$ we just need to apply $\Ibb^{\otimes (n-2)} \otimes X\otimes X $ to the state $\ket{0}^{\log(n))}$. So in general, we just need a depth-1 quantum circuit to construct arbitrary computational basis from $\ket{0}$. Then by lemma \ref{lemma: improveddme}, it takes just a $\mathcal{O}(1)$-depth quantum circuit plus $\log(n)$ ancilla qubits to construct the (exact) block encoding of the so-called projector $\ket{j}\bra{j}$. Let $U_j$ denote the unitary block encoding of the projector $\ket{j}\bra{j}$. We also remark on the following two things. First:
\begin{align}
    \ket{j}\bra{j} \cdot  \textbf{X} = x_{j} \ket{j}\bra{j}
\end{align}
It means that with lemma \ref{lemma: product}, we can transform the block encoding of $\textbf{X}$ into the block encoding of $x_{j} \ket{j}\bra{j} $. Second, given any $\ket{j}$ and $\ket{k}$, it is simply to construct a unitary, denoted as $U(j \rightarrow k)$, composing solely of $X$ gates, that transform $\ket{j}$ to $\ket{k}$. As $U(j \rightarrow k)$ is purely unitary, it encodes itself, and we can use any of the tools provided in the Appendix \ref{sec: prelim}. So, using lemma \ref{lemma: product}, we can construct the block encoding of:
\begin{align}
    U(j\rightarrow k) \big( x_{j} \ket{j}\bra{j} \big) U^\dagger (j \rightarrow k) = x_{j} \ket{k}\bra{k}
\end{align}
In matrix form, the operator $x_{j} \ket{k}\bra{k} $ is a diagonal matrix that contains $x_{j}$ in the $k$-th entry along the diagonal. Hence, the aforementioned procedure essentially allows us, with input a unitary block encoding of some diagonal matrix, to output another diagonal matrix with only one non-zero entry in any location of interest (along the diagonal). The complexity of the procedure is mainly dominated by the use of Lemma \ref{lemma: diagonal} and Lemma \ref{lemma: product}, which therefore involves a $\mathcal{O}(\log(n))$-depth quantum circuit plus $\log(n)+3$ ancillary qubits. We recapitulate the aforementioned procedure in the following lemma.
\begin{lemma}
    Given a unitary block encoding of a diagonal matrix $n \times n$ $\textbf{X} = \sum_{j=1}^{n} x_j \ket{j-1}\bra{j-1}$ of depth $\mathcal{O}(\log(n))$, then given two integers $1\leq j,k \leq n$ there exists a quantum circuit of depth $\mathcal{O}(\log(n))$ (using block encoding of $\textbf{X}$ two times), and $\log(n) +3$ ancilla qubits that produces the block encoding of a matrix of 1 entry $x_j \ket{k}\bra{k}$.
    \label{lemma: singleentry}
\end{lemma}

To exemplify the above lemma, suppose we have a unitary block encoding of the operator:
\begin{align}
    \begin{pmatrix}
        x_1 & 0 & 0 & 0 \\
        0  & x_2 & 0 & 0 \\
        0 & 0  & x_3 & 0\\
        0 & 0 & 0 & x_4
    \end{pmatrix}
\end{align}
By using the above procedure, we can obtain a block encoding of the following operators: 
\begin{align}
    \begin{pmatrix}
        x_1 & 0 & 0 & 0 \\
        0  & 0 & 0 & 0 \\
        0 & 0  & 0 & 0\\
        0 & 0 & 0 & 0
    \end{pmatrix} , 
    \begin{pmatrix}
        0 & 0 & 0 & 0 \\
        0  & x_1 & 0 & 0 \\
        0 & 0  & 0 & 0\\
        0 & 0 & 0 & 0
    \end{pmatrix}, 
    \begin{pmatrix}
        0 & 0 & 0 & 0 \\
        0  & 0 & 0 & 0 \\
        0 & 0  & x_1 & 0\\
        0 & 0 & 0 & 0
    \end{pmatrix},
    \begin{pmatrix}
        0 & 0 & 0 & 0 \\
        0  & 0 & 0 & 0 \\
        0 & 0  & 0 & 0\\
        0 & 0 & 0 & x_1
    \end{pmatrix} \\
    \begin{pmatrix}
        x_2 & 0 & 0 & 0 \\
        0  & 0 & 0 & 0 \\
        0 & 0  & 0 & 0\\
        0 & 0 & 0 & 0 
    \end{pmatrix}, 
    \begin{pmatrix}
        0 & 0 & 0 & 0 \\
        0  & x_2 & 0 & 0 \\
        0 & 0  & 0  & 0\\
        0 & 0 & 0 & 0
    \end{pmatrix},
    \begin{pmatrix}
        0 & 0 & 0 & 0 \\
        0  & 0 & 0 & 0 \\
        0 & 0  & x_2 & 0\\
        0 & 0 & 0 & 0
    \end{pmatrix},
    \begin{pmatrix}
        0 & 0 & 0 & 0 \\
        0  & 0 & 0 & 0 \\
        0 & 0  & 0 & 0\\
        0 & 0 & 0 & x_2
    \end{pmatrix} \\
    \begin{pmatrix}
        x_3 & 0 & 0 & 0 \\
        0  & 0 & 0 & 0 \\
        0 & 0  & 0 & 0\\
        0 & 0 & 0 & 0
    \end{pmatrix} , 
    \begin{pmatrix}
        0 & 0 & 0 & 0 \\
        0  & x_3 & 0 & 0 \\
        0 & 0  & 0 & 0\\
        0 & 0 & 0 & 0
    \end{pmatrix}, 
    \begin{pmatrix}
        0 & 0 & 0 & 0 \\
        0  & 0 & 0 & 0 \\
        0 & 0  & x_3 & 0\\
        0 & 0 & 0 & 0
    \end{pmatrix},
    \begin{pmatrix}
        0 & 0 & 0 & 0 \\
        0  & 0 & 0 & 0 \\
        0 & 0  & 0 & 0\\
        0 & 0 & 0 & x_3
    \end{pmatrix} \\
    \begin{pmatrix}
        x_4 & 0 & 0 & 0 \\
        0  & 0 & 0 & 0 \\
        0 & 0  & 0 & 0\\
        0 & 0 & 0 & 0
    \end{pmatrix} , 
    \begin{pmatrix}
        0 & 0 & 0 & 0 \\
        0  & x_4 & 0 & 0 \\
        0 & 0  & 0 & 0\\
        0 & 0 & 0 & 0
    \end{pmatrix}, 
    \begin{pmatrix}
        0 & 0 & 0 & 0 \\
        0  & 0 & 0 & 0 \\
        0 & 0  & x_4 & 0\\
        0 & 0 & 0 & 0
    \end{pmatrix},
    \begin{pmatrix}
        0 & 0 & 0 & 0 \\
        0  & 0 & 0 & 0 \\
        0 & 0  & 0 & 0\\
        0 & 0 & 0 & x_4
    \end{pmatrix} 
\end{align}

Now we proceed to describe our quantum gradient descent algorithm by discussing the gradient property of the function $f$ that we have assumed. Recall that our function of interest is:
\begin{align}
    f(x_1,x_2,...,x_n) &= \sum_{i=1}^K a_i f_i(x_1,x_2, ..., x_n)  \\
    &= \sum_{i=1}^K a_i x_1^{i_1} x_2^{i_2} ... x_n^{i_n}
\end{align}
where $i_1,i_2,...,i_n \in \mathbb{Z}$. It is easy to see analytically that, for any $1 \leq m \leq n$:
\begin{align}
\label{16}
    \frac{\partial  f(x_1,x_2,...,x_n)}{\partial x_m} =  \sum_{i=1}^K a_i \frac{ \partial f_i(x_1,...,x_n) }{\partial x_m} \begin{cases}
        = \sum_i a_i i_m x_1^{i_1} x_2^{i_2} ... x_m^{ (i_m-1) } ... x_n^{i_n} \text{ if $i_m \geq 1$}\\
        = 0 \text{ if $i_m = 0$}
    \end{cases}
\end{align}
which means that essentially, the derivative of function $f$ with respect to any variable is again a sum of monomial. Our goal now is to construct such gradient from the tool we developed in lemma \ref{lemma: singleentry}. We first consider $f_1(x_1,x_2,...,x_n)$. Its gradient is a column vector, and we embedded such column vector in a matrix:
\begin{align}
   \bigtriangledown f_1(x_1,x_2,...,x_n) = \begin{pmatrix}
        \frac{\partial  f_1(x_1,x_2,...,x_n)}{\partial x_1} \\
        \frac{\partial  f_1(x_1,x_2,...,x_n)}{\partial x_2}\\
        \vdots \\
        \frac{\partial  f_1(x_1,x_2,...,x_n)}{\partial x_n} 
    \end{pmatrix} 
    \longrightarrow 
    \begin{pmatrix}
        \frac{\partial  f_1(x_1,x_2,...,x_n)}{\partial x_1} & 0 & 0 & 0 \\
        0 & \frac{\partial  f_1(x_1,x_2,...,x_n)}{\partial x_2} & 0 & 0 \\
        0 & 0 & \ddots & 0 \\
        0 & 0 & 0 & \frac{\partial  f_1(x_1,x_2,...,x_n)}{\partial x_n} 
    \end{pmatrix}
\end{align}
As we set the convention, $v(f_1)$ refers to the number of variables of $f_1$, and $s(f_1)$ refers to the set of these variables. Therefore, its gradient only contains $v(f_1) = |s(f_1)|$ non-zero entries, which means that along the diagonal of the matrix on right-hand side above, only $v(f_1)$ entries are non-zero. Without loss of generality, suppose that, the partial derivative of $f_1$ with respect to $x_1$ is one of non-zero ones, i.e., $x_1 \in s(f_1)$. Then we have that: 
\begin{align}
    \frac{\partial f_1(x_1,x_2,...,x_n)}{ \partial x_1} = a_1 i_1 x_1^{i_1-1} x_2^{i_2} ... x_n^{i_n}
\end{align}
We quickly remind a subtle detail that out of the $n$ values $(i_1,i_2,...,i_n)$, there are $n-v(f_1)$ terms are zero. Additionally, we from the lemma \ref{lemma: singleentry}, we have block encoding of the following $n\times n$ operators:
\begin{align}
\begin{pmatrix}
    x_1 & 0 & \cdots & 0 \\
    0 & 0 & \cdots & 0 \\
    0 & 0 & \ddots & 0 \\
    0 & 0 & \cdots & 0 
\end{pmatrix}, \begin{pmatrix}
    x_2 & 0 & \cdots & 0 \\
    0 & 0 & \cdots & 0 \\
    0 & 0 & \ddots & 0 \\
    0 & 0 & \cdots & 0 
\end{pmatrix}, \begin{pmatrix}
     x_3 & 0 & \cdots & 0 \\
    0 & 0 & \cdots & 0 \\
    0 & 0 & \ddots & 0 \\
    0 & 0 & \cdots& 0 
\end{pmatrix}, \cdots , \begin{pmatrix}
     x_n & 0 & \cdots & 0 \\
    0 & 0 & \cdots& 0 \\
    0 & 0 & \ddots & 0 \\
    0 & 0 &\cdots & 0 
\end{pmatrix}
\end{align}
where we use the symbol $(.)$ to simply denote the zero entries. Now we outline a simple produce that  construct the term $ a_1 i_1 x_1^{i_1-1} x_2^{i_2} ... x_n^{i_n}$ embedded in the first entry of the $n\times n$ matrix, e.g., $a_1 i_1 x_1^{i_1-1} x_2^{i_2} ... x_n^{i_n} \ket{0}\bra{0}$. 

\begin{method}
\label{algo: firstderivative}
\end{method}
Suppose that we have the block encoding of the following operators $\{ x_m \ket{0}\bra{0} \} $ for $m=1,2,...,n$.
\begin{enumerate}
       \item First, for each block encoding of the operator $ x_m \ket{0}\bra{0}$ (for $m=1,2,...,n$) such that $i_m \geq 1$, we can use lemma \ref{lemma: product} to construct the block encoding of $x_m^{i_m} \ket{0} \bra{0}$ for each $m=2,...,n$. If $i_m =0$, then we simply ignore it and only proceed with those operators with $i_m >0$. For $m=1$, we construct the block encoding of $x_1^{i_1-1} \ket{0}\bra{0}$ instead. If $i_1 =1$, then the block encoding of operator $\ket{0}\bra{0}$ is trivially obtained.
       \item Use lemma \ref{lemma: product} again to construct the block encoding of $x_1^{i_1-1} x_2^{i_2} ... x_n^{i_n} \ket{0}\bra{0}$. 
       \item Use lemma \ref{lemma: scale} to construct the block encoding of $a_1 x_1^{i_1-1} x_2^{i_2} ... x_n^{i_n} \ket{0}\bra{0}$. 
       \item Use lemma \ref{lemma: scale} again to construct the block encoding of $a_1 \frac{i_1}{M} x_1^{i_1-1} x_2^{i_2} ... x_n^{i_n} \ket{0}\bra{0}$. 
\end{enumerate}
   where we remind that in the last line, $M$ is the upper bound of the norm of gradient of $f$ within the domain $\mathscr{D}$. 
According to Eqn.~\ref{16}, the output of the above algorithm is $ \frac{1}{M} \frac{\partial f_1(x_1,...,x_n)}{ \partial x_1} \ket{0}\bra{0}$, which is assumed to be non-zero among $v(f_1)$ non-zero gradient entries. Now we consider another variable $x_p \in s(f_1)$ and that the partial derivative of $f_1$ with respect to $x_p$ is also non-zero 
$$  \frac{\partial f_1(x_1,...,x_n)}{ x_p } = i_p x_1^{i_1} x_2^{i_2} ... x_p^{i_p-1} ... x_m x_n^{i_n} $$
It is straightforward to extend and apply the above algorithm \ref{algo: firstderivative} with the following input $\{ x_m \ket{p-1}\bra{p-1} \}_{m=1}^n$ (which could be obtained from lemma \ref{lemma: singleentry}) to construct the block encoding of the following operator:
\begin{align}
    \frac{i_p}{M} x_1^{i_1} x_2^{i_2} ... x_p^{i_p-1} ... x_n^{i_n} \ket{p-1} \bra{p-1} \equiv \frac{1}{M} \frac{\partial f_1(x_1,x_2,...,x_n )}{ \partial x_p} \ket{p-1}\bra{p-1}
\end{align}
Hence, we can repeat the same procedure and obtain all the non-zero partial derivatives. Lemma \ref{lemma: sumencoding} then can be applied to obtain the block encoding of their summation:
\begin{align}
    \frac{1}{v(f_i) M} \sum_{x_j \in s(f_1)}^ \frac{1}{M} \frac{\partial f_1(x_1,x_2,...,x_n) }{\partial x_j} \ket{j-1}\bra{j-1} 
\end{align}
Given the fact that for all variables $\notin s(f_1)$, the partial derivative of $f_1$ with respect to those variables are zero. So the above summation is in fact:
\begin{align}
    U_1 \equiv \frac{1}{v(f_1) M} \begin{pmatrix}
        \frac{\partial  f_1(x_1,x_2,...,x_n)}{\partial x_1} & 0 & 0 & 0 \\
        0 & \frac{\partial  f_1(x_1,x_2,...,x_n)}{\partial x_2} & 0 & 0 \\
        0 & 0 & \ddots & 0 \\
        0 & 0 & 0 & \frac{\partial  f_1(x_1,x_2,...,x_n)}{\partial x_n} 
    \end{pmatrix}
\end{align}

Then the above procedure can be repeated, but with another function $f_2(x_1,x_2,...,x_n)$. Then we obtain the block encoding of:
\begin{align}
    & U_2 \equiv \frac{1}{v(f_2) M} \begin{pmatrix}
        \frac{\partial  f_2(x_1,x_2,...,x_n)}{\partial x_1} & 0 & 0 & 0 \\
        0 & \frac{\partial  f_2(x_1,x_2,...,x_n)}{\partial x_2} & 0 & 0 \\
        0 & 0 & \ddots & 0 \\
        0 & 0 & 0 & \frac{\partial  f_2(x_1,x_2,...,x_n)}{\partial x_n} 
    \end{pmatrix}, \\
    & \vdots \\
    & U_K \equiv \frac{1}{v(f_K) M} \begin{pmatrix}
        \frac{\partial  f_K(x_1,x_2,...,x_n)}{\partial x_1} & 0 & 0 & 0 \\
        0 & \frac{\partial  f_K(x_1,x_2,...,x_n)}{\partial x_2} & 0 & 0 \\
        0 & 0 & \ddots & 0 \\
        0 & 0 & 0 & \frac{\partial  f_K(x_1,x_2,...,x_n)}{\partial x_n} 
    \end{pmatrix}
\end{align}
Now we can use the amplification tool Lemma \ref{lemma: amp_amp} to (approximately) remove the factor $v(f_i)$ (for $i=1,2,...,K$) from the above operators, e.g.:
\begin{align}
    U_1 \longrightarrow \frac{1}{2} v(f_1) U_1, U_2 \longrightarrow \frac{1}{2}v(f_2) U_2, \cdots, U_K \longrightarrow \frac{1}{2} v(f_K) U_K
\end{align}
where each amplification requires a quantum circuit of depth $\mathcal{O}( v(f_i) \log (v(f_i)/\epsilon) )$ where $\epsilon$ is the error tolerance. We give a quick explanation for why the above is possible, as well as the factor $1/2$ and $\epsilon$ in the complexity. The reason for the factor $1/2$ above is that, when we apply lemma \ref{lemma: amp_amp}, we choose the parameter $\delta$ in such lemma to be $1/2$, and it is required that the norm of the resulting operator, for example, $ \frac{1}{2} v(f_1) U_1$ must be less than $1/2$. We further note that:
\begin{align}
    \frac{1}{2} v(f_1) U_1 = \frac{1}{2 M} \begin{pmatrix}
        \frac{\partial  f_1(x_1,x_2,...,x_n)}{\partial x_1} & 0 & 0 & 0 \\
        0 & \frac{\partial  f_1(x_1,x_2,...,x_n)}{\partial x_2} & 0 & 0 \\
        0 & 0 & \ddots & 0 \\
        0 & 0 & 0 & \frac{\partial  f_1(x_1,x_2,...,x_n)}{\partial x_n} 
    \end{pmatrix}
\end{align}
As we assumed that the norm of gradient of $f$, $|\bigtriangledown f(x_1,x_2,...,x_n)|$ is less than $M$ for any $(x_1,...,x_n) \in \mathscr{D}$, which means that: 
\begin{align}
   \sqrt{ \sum_{j=1}^n ( \frac{\sum_{i=1}^K \partial f_i(x_1,x_2,...,x_n)}{\partial x_j} )^2 } \leq M
\end{align}
which suggests that for any $i,j$, $| \partial f_i(x_1,...,x_n)/\partial x_j| \leq M$, which implies that the matrix norm $| \frac{1}{2} v(f_1) U_1 | = \max_j \frac{1}{2M} |  \frac{\partial f_1(x_1,...,x_n)}{\partial x_j} | \leq 1/2$. Hence, it satisfies the requirement for applying lemma \ref{lemma: amp_amp}. Regarding the error factor $\epsilon$, it simply means that once we apply the amplification tool, the result is an $\epsilon$-approximated (see definition \ref{def: blockencode}) block encoding of corresponding operator, e.g., $\frac{1}{2} v(f_1) U_1$. As a next step, we can use lemma \ref{lemma: sumencoding} to construct the block encoding of the following operator: 
\begin{align}
    & \frac{1}{K} \sum_{i=1}^K \big(\frac{1}{2}  v(f_1) U_1 + \frac{1}{2}v(f_2) U_2 + ... +\frac{1}{2} v(f_K) U_K \big)  \\
    &=  \frac{1}{2MK} \sum_{i=1}^K \begin{pmatrix}
        \frac{\partial  f_i(x_1,x_2,...,x_n)}{\partial x_1} & 0 & 0 & 0 \\
        0 & \frac{\partial  f_i(x_1,x_2,...,x_n)}{\partial x_2} & 0 & 0 \\
        0 & 0 & \ddots & 0 \\
        0 & 0 & 0 & \frac{\partial  f_i(x_1,x_2,...,x_n)}{\partial x_n} 
    \end{pmatrix} \\
    &= \frac{1}{2MK} \begin{pmatrix}
        \frac{\partial  f(x_1,x_2,...,x_n)}{\partial x_1} & 0 & 0 & 0 \\
        0 & \frac{\partial  f(x_1,x_2,...,x_n)}{\partial x_2} & 0 & 0 \\
        0 & 0 & \ddots & 0 \\
        0 & 0 & 0 & \frac{\partial  f(x_1,x_2,...,x_n)}{\partial x_n} 
    \end{pmatrix}
\end{align}
We note that from the previous step, we have an $\epsilon$-approximated block encoding of operator $(1/2) v(f_1)U_1, ...., (1/2) v(f_n) U_n$, and hence, the resulting block encoding of the above operator is also $\epsilon$-approximated. We recapitulate the aforementioned procedure in the following lemma:
\begin{lemma}[Construction of gradient]
    \label{lemma: constructiongradient} 
    Let $f: \Rbb^n \longrightarrow \Rbb $ be some $n$-variables function and that $f$ could be decomposed as $f = \sum_{i=1}^K f_i$ where each $f_i$ is a monomial of $v(f_i)$ variables with total degree $d(f_i)$. Let $v = \max_{i=1}^K v(f_i)$ and $d = \max_{i=1}^K d(f_i)$. Given some $\log(n)$-qubit unitary that generates a state $\sum_{j=1} x_j \ket{j-1}$. Then there exists a quantum produce, with $\mathcal{O}(\log(n) + \log(v) + \log(K)) $ ancilla qubits, and a quantum circuit of depth $\mathcal{O}(\log(n) K^2  d \cdot v^2  \log(\frac{1}{\epsilon}))$, that outputs an $\epsilon$-approximated block encoding of the operator
    $$ \frac{1}{2MK} \sum_{i=1}^K \frac{\partial f}{\partial x_i} \ket{i-1}\bra{i-1}$$
\end{lemma}

Recall that gradient descent begins with a random guess $\xbf_0 = (x_{1,0},x_{2,0},...,x_{n,0})$ where we remark that the extra subscript $0$ is used to denote the 0-th iteration step, e.g., $\xbf_T =(x_{1,T},x_{2,T},...,x_{n,T}) $ refers to the temporal solution at $T$-th iteration step. Let $U_0$ be an arbitrary $\log(n)$-qubits unitary, and we set that:
\begin{align}
    U_0 \ket{0}^{\log (n)} = \sum_{j=1}^n x_{j,0} \ket{j-1} 
\end{align}
Now we begin to formally describe our quantum gradient descent and its complexity.
\begin{method}[Quantum Gradient Descent]
\label{method: quantumgradientdescent}
\end{method}
    Let $f: \Rbb^n \longrightarrow \Rbb$ be as defined earlier, and $U_0$ as above. Assume that $U_0$ has $\mathcal{O}(1)$ depth.
    \begin{enumerate}
        \item Use $U_0$ and lemma \ref{lemma: diagonal} to construct the block encoding of 
        $$ \begin{pmatrix}
        x_{1,0} & 0 & \cdots  & 0 \\
        0 & x_{2,0} & \cdots & 0\\
        0 & 0 & \ddots & 0 \\
        0 & 0 & \cdots & x_{n,0} 
    \end{pmatrix} $$
    The time complexity, or quantum circuit depth of this step is $\mathcal{O}(\log(n))$ and spatial complexity is $\mathcal{O}(\log(n))$
    \item With the above (block encoding of) operator as input, use algorithm \ref{algo: firstderivative} and the procedure from lemma \ref{lemma: constructiongradient} with the above input to construct the $\epsilon$-approximated block encoding of: 
    $$ \frac{1}{2MK} \begin{pmatrix}
        \frac{\partial  f(x_{1,0},x_{2,0},...,x_{n,0})}{\partial x_{1,0}} & 0 & 0 & 0 \\
        0 & \frac{\partial  f(x_{1,0},x_{2,0},...,x_{n,0})}{\partial x_{2,0}} & 0 & 0 \\
        0 & 0 & \ddots & 0 \\
        0 & 0 & 0 & \frac{\partial  f(x_{1,0},x_{2,0},...,x_{n,0})}{\partial x_{n,0}} 
    \end{pmatrix}$$
    The time complexity of this step, including the previous step, is $\mathcal{O}( \log(n)K^2 d v^2 \log(1/\epsilon)  )$ and spatial complexity is $\mathcal{O}(\log(n) + \log(v) + \log(K))$. 
    \item Use lemma \ref{lemma: sumencoding} with minus sign instead of plus sign, to construct the $\epsilon$-approximated block encoding of
    $$ \frac{1}{2 } \big( \begin{pmatrix}
           x_{1,0}  & 0 & \cdots  & 0 \\
        0 & x_{2,0} & \cdots & 0\\
        0 & 0 & \ddots & 0 \\
        0 & 0 & \cdots & x_{n,0} 
    \end{pmatrix} - \frac{1}{2MK} \begin{pmatrix}
        \frac{\partial  f(x_{1,0},x_{2,0},...,x_{n,0})}{\partial x_{1,0}} & 0 & 0 & 0 \\
        0 & \frac{\partial  f(x_{1,0},x_{2,0},...,x_{n,0})}{\partial x_{2,0}} & 0 & 0 \\
        0 & 0 & \ddots & 0 \\
        0 & 0 & 0 & \frac{\partial  f(x_{1,0},x_{2,0},...,x_{n,0})}{\partial x_{n,0}} 
    \end{pmatrix} \big)$$
     Treat $1/(2MK) \equiv \eta$ as the hyperparameter in the gradient descent, we have that the above operator is essentially 
    \begin{align*}
     &\frac{1}{2 } \big( \begin{pmatrix}
           x_{1,0}  & 0 & \cdots  & 0 \\
        0 & x_{2,0} & \cdots & 0\\
        0 & 0 & \ddots & 0 \\
        0 & 0 & \cdots & x_{n,0} 
    \end{pmatrix} - \eta \begin{pmatrix}
        \frac{\partial  f(x_{1,0},x_{2,0},...,x_{n,0})}{\partial x_{1,0}} & 0 & 0 & 0 \\
        0 & \frac{\partial  f(x_{1,0},x_{2,0},...,x_{n,0})}{\partial x_{2,0}} & 0 & 0 \\
        0 & 0 & \ddots & 0 \\
        0 & 0 & 0 & \frac{\partial  f(x_{1,0},x_{2,0},...,x_{n,0})}{\partial x_{n,0}} 
    \end{pmatrix} \big) \\ 
    &= \frac{1}{2} \begin{pmatrix}
           x_{1,1}  & 0 & \cdots  & 0 \\
        0 & x_{2,1} & \cdots & 0\\
        0 & 0 & \ddots & 0 \\
        0 & 0 & \cdots & x_{n,1} 
    \end{pmatrix}
     \end{align*}
     where we remind that the new subscript $1$ has been used, indicating that the solution is updated after the first gradient descent step. The above factor $1/2$ can be removed by using amplification \ref{lemma: amp_amp}, resulting the $\epsilon$-closed approximated of 
     $$ \begin{pmatrix}
           x_{1,1}  & 0 & \cdots  & 0 \\
        0 & x_{2,1} & \cdots & 0\\
        0 & 0 & \ddots & 0 \\
        0 & 0 & \cdots & x_{n,1} 
    \end{pmatrix} $$
     Thus, the time complexity of this step (including previous two steps) is $\mathcal{O}( \log(n)K^2 d v^2 \log(\frac{1}{\epsilon}) )$ and spatial complexity is $\mathcal{O}(\log(n) + \log(v) + \log(K))$.
     \item Iterate the following procedure $T$ times: use the $\epsilon$-approximated block encoding of the above operator and repeat from step (2). Eventually, we obtain an $\epsilon$-approximated block encoding of 
     $$ \begin{pmatrix}
           x_{1,T}  & 0 & \cdots  & 0 \\
        0 & x_{2,T} & \cdots & 0\\
        0 & 0 & \ddots & 0 \\
        0 & 0 & \cdots & x_{n,T} 
    \end{pmatrix}$$
    The total time complexity, or circuit depth of the algorithm is $ \mathcal{O}\Big( \log(n) \big(K^2 d v^2 \log\frac{1}{\epsilon} \big)^T \Big)$ and the spatial complexity is $\mathcal{O}\big(\log(n) + \log(v) + \log(K)\big) $. 
    \end{enumerate}

\subsection{Less generic but potentially more practical type of function}
\label{sec: anothertype}
The above algorithm, despite being applicable to a generic type of function, turns out to have scaling $\sim (Kv)^T$ where $K$ is the number of monomials. Thus, it is only effective when $v,K$ is small. Consider the following three-variable function $f(x_1, x_2, x_3) = g(x_1) + h(x_2) + p(x_3)$ where $g,h,p$ are some functions in respective variables $x_1,x_2,x_3$. As the number of terms is $K = 3$, the algorithm above would have scaling at least 3, which is also the number of variables, and hence implying that there is no exponential speed-up. More generally, if $K = \mathcal{O}(n)$, then the logarithmical scaling in the number of variables $n$ (of Algorithm.~\ref{method: quantumgradientdescent}) is not going to be useful. Given the power of QSVT especially for handling polynomials, it is reasonable to believe that the capability of quantum computer should go beyond the limit of small $K$. In the following, we impose a certain structure on the function $f(x_1,x_2,..,x_n)$ and provide a quantum gradient descent algorithm based on that. Despite being less generic than the previous case, we envision that it will be more practically useful. 

Consider a multi-variate function $f: \Rbb^n \longrightarrow \Rbb$ of the form 
\begin{align}
    f(x_1,x_2,...,x_n) =  \sum_{i=1}^n \mathscr{F}(x_i) = \mathscr{F}(x_1) + \mathscr{F}(x_2) + ... + \mathscr{F}(x_n)
\end{align}
where $\mathscr{F}(x_i)$ is some analytical function of corresponding variable. Similar to previous case, we assume that our domain of interest is $\mathscr{D} = [-1/2,1/2]^n$ and $|f(x_1,x_2,...,x_n)| \leq 1/2$ in this domain $\mathscr{D}$. Additionally, the norm of gradient of $f$ is bounded by some known value $M$, i.e., $| \bigtriangledown f(x_1,x_2,...,x_n) | \leq M $ for all $(x_1,x_2,...,x_n) \in \mathscr{D}$. It is obvious that, for any $i$:
\begin{align}
    \frac{\partial f(x_1,x_2,...,x_n)}{\partial x_i} = \frac{\partial \mathscr{F}(x_i)}{\partial x_i}
\end{align}
Hence the gradient is:
\begin{align}
    \bigtriangledown f(x_1,x_2,...,x_n) = \big( \frac{\partial \mathscr{F}(x_1)}{\partial x_1}, \frac{\partial \mathscr{F}(x_2)}{\partial x_2}, ..., \frac{\partial \mathscr{F}(x_n)}{\partial x_n} \big)^T
\end{align}
As long as $\mathscr{F}(x_i)$ is an analytical function, its derivative admits an analytical expression. Assuming that for all $i$, $ \frac{\partial \mathscr{F}(x_i)}{\partial x_i}$ can be approximated by a polynomial of a certain degree, which we denote by $\mathcal{P}$. In principle, if $\mathscr{F}(x_i)$ admits a polynomial approximation, then its derivative also admits. We point out that in \cite{gilyen2019quantum}, it is shown that a wide range of functions can be well approximated by polynomial. By being well approximated, it means that, to approximate the function to additive error $\epsilon$, a polynomial of degree $\mathcal{O}\big(\log \frac{1}{\epsilon}\big)$ is sufficient. This assumption allows us to leverage the following central tool from \cite{gilyen2019quantum}:
\begin{lemma}\label{lemma: qsvt}[\cite{gilyen2019quantum} Theorem 56]
\label{lemma: theorem56}  \revise{\bf (Polynomial Transformation)}
Suppose that $U$ is an
$(\alpha, a, \epsilon)$-encoding of a Hermitian matrix $A$. (See Definition 43 of~\cite{gilyen2019quantum} for the definition.)
If $P \in \mathbb{R}[x]$ is a degree-$d$ polynomial satisfying that
\begin{itemize}
\item for all $x \in [-1,1]$: $|P(x)| \leq \frac{1}{2}$,
\end{itemize}
then, there is a quantum circuit $\tilde{U}$, which is an $(1,a+2,4d \sqrt{\frac{\epsilon}{\alpha}})$-encoding of $P(A/\alpha)$ and
consists of $d$ applications of $U$ and $U^\dagger$ gates, a single application of controlled-$U$ and $\mathcal{O}((a+1)d)$
other one- and two-qubit gates.
\end{lemma}
We have enough recipes to describe our quantum gradient descent algorithm. In the following, we formally outline it. 
\begin{method}
    \label{method: anotherquantumalgorithm}
\end{method}
Let $f:\Rbb^n \longrightarrow \Rbb$ be defined as earlier, and a unitary $U_0$ that generates an initial state $\xbf_{0}$ as $U_0 \ket{0}^{\log (n)} = \sum_{j=1}^n x_{j,0} \ket{j-1} $.
\begin{enumerate}
    \item Use $U_0$ and Lemma \ref{lemma: diagonal} to construct the block encoding of 
    $$\begin{pmatrix}
        x_{1,0} & 0 & \cdots  & 0 \\
        0 & x_{2,0} & \cdots & 0\\
        0 & 0 & \ddots & 0 \\
        0 & 0 & \cdots & x_{n,0} 
    \end{pmatrix} $$
    The time complexity, or quantum circuit depth of this step is $\mathcal{O}(\log(n))$ and spatial complexity is $\mathcal{O}(\log(n))$
    \item Use Lemma \ref{lemma: theorem56} and transform the above block-encoded operator into:
    \begin{align}
        \begin{pmatrix}
        x_{1,0} & 0 & \cdots  & 0 \\
        0 & x_{2,0} & \cdots & 0\\
        0 & 0 & \ddots & 0 \\
        0 & 0 & \cdots & x_{n,0} 
    \end{pmatrix} \longrightarrow  \frac{1}{M}\begin{pmatrix}
       \mathcal{P}\big(x_{1,0}\big) & 0 & \cdots  & 0 \\
        0 & \mathcal{P}\big(x_{2,0}\big) & \cdots & 0\\
        0 & 0 & \ddots & 0 \\
        0 & 0 & \cdots & \mathcal{P}\big(x_{n,0} \big)
    \end{pmatrix}
    \end{align}
    The complexity of this step is $\mathcal{O}\big( \deg(\mathcal{P}) \log(n) \big)$
    \item Use either Lemma \ref{lemma: amp_amp}, or Lemma \ref{lemma: scale} to insert the hyperparameter $\eta$ to the above operator, i.e., we obtain the transformation: 
    \begin{align}
        \frac{1}{M}\begin{pmatrix}
       \mathcal{P}\big(x_{1,0}\big) & 0 & \cdots  & 0 \\
        0 & \mathcal{P}\big(x_{2,0}\big) & \cdots & 0\\
        0 & 0 & \ddots & 0 \\
        0 & 0 & \cdots & \mathcal{P}\big(x_{n,0} \big)
    \end{pmatrix} \longrightarrow \eta \begin{pmatrix}
       \mathcal{P}\big(x_{1,0}\big) & 0 & \cdots  & 0 \\
        0 & \mathcal{P}\big(x_{2,0}\big) & \cdots & 0\\
        0 & 0 & \ddots & 0 \\
        0 & 0 & \cdots & \mathcal{P}\big(x_{n,0} \big)
    \end{pmatrix}
    \end{align}
    The complexity of this step is $\mathcal{O}\big( \deg(\mathcal{P}) \log(n) \big)$
    \item Use Lemma \ref{lemma: sumencoding} to construct the block encoding of:
    \begin{align}
       \frac{1}{2} \left( \begin{pmatrix}
        x_{1,0} & 0 & \cdots  & 0 \\
        0 & x_{2,0} & \cdots & 0\\
        0 & 0 & \ddots & 0 \\
        0 & 0 & \cdots & x_{n,0} 
    \end{pmatrix}  -  \eta\begin{pmatrix}
       \mathcal{P}\big(x_{1,0}\big) & 0 & \cdots  & 0 \\
        0 & \mathcal{P}\big(x_{2,0}\big) & \cdots & 0\\
        0 & 0 & \ddots & 0 \\
        0 & 0 & \cdots & \mathcal{P}\big(x_{n,0} \big)
    \end{pmatrix}  \right) = \frac{1}{2} \begin{pmatrix}
           x_{1,1}  & 0 & \cdots  & 0 \\
        0 & x_{2,1} & \cdots & 0\\
        0 & 0 & \ddots & 0 \\
        0 & 0 & \cdots & x_{n,1} 
    \end{pmatrix} 
    \end{align}
    The complexity of this step is $\mathcal{O}\big( \deg(\mathcal{P}) \log(n) \big)$. 
    \item Use Lemma \ref{lemma: amp_amp} and remove the factor $1/2$ in the above operator. Then we obtain an $\epsilon$-approximated block encoding of:
    $$  \begin{pmatrix}
           x_{1,1}  & 0 & \cdots  & 0 \\
        0 & x_{2,1} & \cdots & 0\\
        0 & 0 & \ddots & 0 \\
        0 & 0 & \cdots & x_{n,1} 
    \end{pmatrix}$$
    The complexity of this step is $\mathcal{O}\big( \deg(\mathcal{P}) \log(n) \log \frac{1}{\epsilon}\big)$
    \item Use the above operator and repeat from the beginning, e.g., Step 1-5, we then obtain an $\epsilon$-approximated block encoding of 
    $$ \begin{pmatrix}
           x_{1,T}  & 0 & \cdots  & 0 \\
        0 & x_{2,T} & \cdots & 0\\
        0 & 0 & \ddots & 0 \\
        0 & 0 & \cdots & x_{n,T} 
    \end{pmatrix}$$
    The total time complexity is $\mathcal{O}\left( \log(n)\Big(  \deg(\mathcal{P})  \log \frac{1}{\epsilon}\Big)^T  \right) $ and spatial complexity is $\mathcal{O}\big( \log n  \big)$
\end{enumerate}

\section{Discussion }
\label{sec: discussion}

\noindent
\textbf{Initial condition. } There is one subtle detail in the above two quantum gradient descent algorithms. At the end of step (3) of Algorithm \ref{method: quantumgradientdescent} and step (4) of Algorithm \ref{method: anotherquantumalgorithm}, there is a requirement for using amplification lemma \ref{lemma: amp_amp} and that we require the operator 
\begin{align}
    \begin{pmatrix}
           x_{1,t}  & 0 & \cdots  & 0 \\
        0 & x_{2,t} & \cdots & 0\\
        0 & 0 & \ddots & 0 \\
        0 & 0 & \cdots & x_{n,t} 
    \end{pmatrix}
\end{align}
to have its matrix norm less than $1/2$ at any time step $t$ among the total $T$ iterations, which is equivalent to $ \max_{i=1}^n \{ |x_i,t| \} \leq 1/2$. So, the question is, how to guarantee such a condition? Recall that the gradient descent (both classical and quantum) begins with an initial vector $ \xbf_0 = (x_{1,0}, x_{2,0}, ... , x_{n,0} )$, and iterate the procedure:
\begin{align}
    \xbf_{t+1} = \xbf_t - \eta \bigtriangledown f(\xbf_t) 
\end{align}
From the above equation, we have that:
\begin{align}
    |\xbf_{t+1}|^2 &= |(\xbf_t - \eta \bigtriangledown f(\xbf_t) )|^2 \\ 
    & \leq  (  |\xbf_t| + \eta | \bigtriangledown f(\xbf_t) | )^2  \\
    & \leq ( |\xbf_t| + \eta M )^2 
\end{align}
It means that after each iteration, the norm square of $\xbf_t$ increases by at most $\eta M$. Therefore, if we begin at some $\xbf_0$ with some norm square $|\xbf_0|^2$, then we have that after $T$ iterations, the norm square $\xbf_T$ is at most $ |\xbf_0| + \eta M T$. We also remark that $|\xbf_{t}|^2 = \sum_{i=1}^n x_{i,t}^2 \geq x_{i,t}^2$ for any $1 \leq i \leq n$, i.e., $\max_{i} |x_{i,t}| \leq |\xbf_{t}|$. If we expect that at all time step $t$ among $T$ iterations, the norm $|\xbf_t|$ is less than $1/2$, then it is sufficient to require $|\xbf_0| + \eta M T  \leq \frac{1}{2} $ thus implying $ |\xbf_0| \leq \frac{1}{2 } -  \eta M T$. Hence, we need to choose the initial guess $\xbf_0$ such that its norm is less than the above known threshold. It is quite simple to achieve such condition, and there should be many ways to do it. Here, we point out a simple way using a depth-1 quantum circuit. Suppose we take $q$-qubit state $\ket{0}^{\otimes q}$, and apply $H^{\otimes q}$ to it, we then obtain:
\begin{align}
    H^{\otimes q} \ket{0}^{\otimes q } = \sum_{k=0}^{2^q-1  } \frac{1}{\sqrt{2^q}} \ket{k}
\end{align}
Then lemma \ref{lemma: diagonal} allows us to construct the block encoding of the operator $ \sum_{k=0}^{2^q -1} \frac{1}{\sqrt{2^q}} \ket{k}\bra{k}$. If we pay only attention to the top-left $n \times n$ matrix of such an operator, then the norm of this $n \times n$ matrix is simply $1/\sqrt{2^q}$. As we demand the norm to be less than $1/2 - \eta MT$, we simply need to have $ \frac{1}{\sqrt{2^q}} = \frac{1}{2} - \eta M T \longrightarrow q = \log_2 \frac{1}{(\frac{1}{2}-\eta MT )^2}$. Therefore, it takes a modest number of qubit, and a depth 1 quantum circuit to construct the initial guess $\xbf_0$ that would guarantee the norm of operator 
\begin{align}
    \begin{pmatrix}
           x_{1,t}  & 0 & \cdots  & 0 \\
        0 & x_{2,t} & \cdots & 0\\
        0 & 0 & \ddots & 0 \\
        0 & 0 & \cdots & x_{n,t} 
    \end{pmatrix}
\end{align}
to be less than $1/2$ at any time step $t$-th of total steps $T$. \\

\noindent
\textbf{Obtaining quantum state corresponding to solutions. } The outcomes of both algorithm \ref{method: quantumgradientdescent} and \ref{method: anotherquantumalgorithm} is a block encoding of a diagonal operator that contains the solution $\xbf_T$. In the following, we discuss a simple way to obtain a quantum state $\ket{\xbf_T}$ that corresponds to $\xbf_T$. Let $U_T$ denotes the unitary block encoding of 
\begin{align}
    \begin{pmatrix}
           x_{1,T}  & 0 & \cdots  & 0 \\
        0 & x_{2,T} & \cdots & 0\\
        0 & 0 & \ddots & 0 \\
        0 & 0 & \cdots & x_{n,T} 
    \end{pmatrix}
\end{align}
According to Definition.~\ref{def: blockencode} and Eqn.~\ref{eqn: action}, we have that:
\begin{align}
    U_T \ket{\bf 0} \frac{1}{\sqrt{n}} \sum_{i=1}^n \ket{i-1} = \ket{\bf 0}  \sum_{i=1}^n \frac{x_{i,T}}{\sqrt{n}} \ket{i-1} + \ket{\rm Garbage}
\end{align}
Performing measurement on the ancilla and post-select on $\ket{\bf 0}$, we obtain the state $\sim \sum_{i=1}^n \frac{x_{i,T}}{\sqrt{n}} \ket{i-1}$, which is exactly $\ket{\xbf_T}$. The success probability is $ \frac{1}{n} \sum_{i=1}^n x_{i,T}^2$. Given that each $x_{i,T} = \mathcal{O}(1)$, it is reasonable to believe that the success probability is $\mathcal{O}(1)$. Given that the state $ \frac{1}{\sqrt{n}} \sum_{i=1}^n \ket{i-1} $  can be prepared with depth-1 circuit, e.g., $H^{\otimes \log n}$, this step incurs only $\mathcal{O}(1)$ complexity. \\

\noindent
\textbf{Comparison to previous work Ref.~\cite{rebentrost2019quantum}.} The above construction yields two quantum gradient descent algorithms that has running time $\mathcal{O}\Big( \log(n) \big(K^2 d v^2 \log\frac{1}{\epsilon} \big)^T \Big)$ and $\mathcal{O}\left( \log(n)\Big( \deg(\mathcal{P}) \log \frac{1}{\epsilon}\Big)^T  \right) $, respectively. The input functions of interest have the form $f(x_1,x_2,...,x_n) = \sum_{i=1}^K a_i x_1^{i_1} x_2^{i_2} ... x_n^{i_n}$, which is a combination of monomials and $f(x_1,x_2,...,x_n) = \mathscr{F}(x_1) + \mathscr{F}(x_2) + ... + \mathscr{F}(x_n) $, which is a sum of polynomials. The first type is more general; however, the complexity is bigger, and as we pointed out earlier, there are some simple cases where it turns out not to be so efficient. The second type, on the other hand, fills that gap. These kinds of structures are more general compared to homogeneous polynomials of even degree that were developed in \cite{rebentrost2019quantum}. More specifically, each term $f_i(x_1,x_2, ..., x_n) $, or $\mathscr{F}(x_i)$, could be of a different degree and is not limited to odd or even ones. 

To compare with \cite{rebentrost2019quantum}, we recall that a homogeneous polynomial of even degree has a tensor algebraic form $f(\xbf) = \frac{1}{2} (\xbf^T)^{\otimes p} A (\xbf)^{\otimes p}$. The time complexity in their work is:
$$\mathcal{O}\Big( \frac{p^{5T} s^T }{ \epsilon^{4T}    } \log(n) \Big) $$
where $s$ is the sparsity of $A$. In the context of Algorithm \ref{method: quantumgradientdescent}, $d = 2p$ and $v=\mathcal{O}(p)$. The total number of terms is $K = \mathcal{O}( s n^p  )$. So our time complexity is:
$$ \mathcal{O} \Big(  \log(n)  (s^2 n^{2p} p^3\log \frac{1}{\epsilon} )^T \Big) $$
Comparing these complexity, we see that our algorithm \ref{method: quantumgradientdescent} does not admit improvement in dimension $n$, however, there is a power-of-two improvement in $p$ and a superpolynomial improvement in the inverse of error tolerance. Thus, our method is more suitable for low-dimensional settings, but with a high degree of polynomial. We point out that the main obstacle that prevents us from achieving a better dependence on $n$ in this case is the scaling term $K^{2T}$, as $K$ turns out to be $\sim sn^p$. In fact, given that $f(\xbf) = \frac{1}{2} (\xbf^T)^{\otimes p} A (\xbf)^{\otimes p}$, there is a possible improvement. We remind the definition of sparsity $s$ is the maximum number of non-zero entries in each row/column. If, for example, among $n^p$ rows of $A$, only $\mathcal{S}$ rows each contains at most $s$ non-zero entries. Then the total of terms in $f(\xbf) = \frac{1}{2} (\xbf^T)^{\otimes p} A (\xbf)^{\otimes p}$ is $K  = s\mathcal{S}$. In this case, termed the case of the highly sparse case, the time complexity is improved to $$ \mathcal{O}\Big( \log(n)  (s^2 \mathcal{S}^2 p\log \frac{1}{\epsilon})^T \Big)$$
If $\mathcal{S}$ is considerably smaller than $n^p$, for example, of order $\mathcal{O}(1), \mathcal{O}\big( \log(n^p)\big)$, then our method can match the complexity scaling of \cite{rebentrost2019quantum} in dimension $n$. The regime of Algorithm \ref{method: anotherquantumalgorithm} turns out not to coincide with the Ref.~\cite{rebentrost2019quantum}. In fact, we have pointed out earlier that there are types of functions that Algorithm \ref{method: anotherquantumalgorithm} can handle more efficiently than Algorithm \ref{method: quantumgradientdescent}. Although the method of~\cite{rebentrost2019quantum} is only applicable to a homogeneous polynomial of even degree, the algorithm \ref{method: anotherquantumalgorithm} introduced above can deal with polynomial of various kinds, and can go even beyond the polynomial domain. Thus, this method is a nice complement to the Ref.~\cite{rebentrost2019quantum} (and also Algorithm \ref{method: quantumgradientdescent}). 

\noindent
\textbf{Potential improvement.} One may wonder how classical computer performs in our setting. As straightforward as it can be, given any $\xbf = (x_1,x_2,...,x_n)$, the classical computer can compute the gradients by taking each partial derivative, resulting in the $\mathcal{O}( n d K v)$ time step. The update step can be performed in another $\mathcal{O}(n)$ time (that is, simply subtracting the entry by the entry). Thus, the total classical time required is $\mathcal{O}( n d K v T)$. So, despite the potential for (poly)logarithmical scaling in dimension $n$, the exponential dependence on $T$ of our method is quite problematic. We remind the readers that $T$ is the total number of iteration steps, and typically it is user-dependent. A few investigation \cite{nesterov1983method,nesterov2013introductory, boyd2004convex} have shown that if the considered function is convex, then $T$ needs to be $\mathcal{O}(\frac{1}{\delta})$ in order to be $\delta$-close to the real point of extrema. If the function is strongly convex, then $T$ is improved to $\mathcal{O}( \log \frac{1}{\delta})$. In this case, our exponential dependence on $T$ will translate into linear dependence on $\frac{1}{\delta}$, which is more efficient.

\section{Conclusion}
\label{sec: conclusion}
While the problem of gradient descent is not entirely new, nor is the quantum singular value transformation (QSVT) framework, our work offers a fresh perspective on both. The most significant advancement of our approach is the elimination of the need for coherent access to any matrix—an assumption in \cite{rebentrost2019quantum} required for applying Hamiltonian simulation techniques. We believe that oracle, or black-box, access significantly limits the practical applicability of quantum algorithms, as the development of such operations, including quantum random access memory, is still in its early stages and remains impractical for large-scale applications.

More broadly, problems involving monomials, polynomials, and related structures can be addressed in a straightforward manner within the quantum framework, thanks to the powerful capabilities of quantum singular value transformation \cite{gilyen2019quantum}. This framework has demonstrated immense potential in unifying the description of various quantum algorithms. However, extending its reach and applicability remains an open and exciting challenge. Our work provides a simple yet compelling example within the optimization domain, serving as strong motivation for further exploration.

We firmly believe that many computational problems exhibit intrinsic structures that can be translated, either directly or indirectly, into a polynomial-based formulation. With the right quantum algorithmic techniques, such problems can be both simplified and improved. For instance, in the original quantum linear system algorithm \cite{harrow2009quantum, childs2017quantum}, the authors assumed black-box access to the matrix before applying Hamiltonian simulation (in a manner similar to \cite{rebentrost2019quantum}) to perform matrix inversion. Inspired by our approach and its concrete applications, alternative representations of matrices may be developed that leverage QSVT, thereby eliminating the need for oracle or black-box access. We conclude our discussion here, leaving this promising direction open for future investigation.

\bibliography{ref.bib}{}
\bibliographystyle{unsrt}

\clearpage
\newpage
\onecolumngrid
\appendix

\section{Preliminaries}
\label{sec: prelim}
Here, we summarize the main recipes of our work, which mostly derived in the seminal QSVT work \cite{gilyen2019quantum}. We keep the statements brief and precise for simplicity, with their proofs/ constructions referred to in their original works.

\begin{definition}[Block Encoding Unitary]~\cite{low2017optimal, low2019hamiltonian, gilyen2019quantum}
\label{def: blockencode} 
Let $A$ be some Hermitian matrix of size $N \times N$ whose matrix norm $|A| < 1$. Let a unitary $U$ have the following form:
\begin{align*}
    U = \begin{pmatrix}
       A & \cdot \\
       \cdot & \cdot \\
    \end{pmatrix}.
\end{align*}
Then $U$ is said to be an exact block encoding of matrix $A$. Equivalently, we can write:
\begin{align*}
    U = \ket{ \bf{0}}\bra{ \bf{0}} \otimes A + \cdots
\end{align*}
where $\ket{\bf 0}$ refers to the ancilla system required for the block encoding purpose. In the case where the $U$ has the form 
$$ U  =  \ket{ \bf{0}}\bra{ \bf{0}} \otimes \Tilde{A} + \cdots $$
where $|| \Tilde{A} - A || \leq \epsilon$ (with $||.||$ being the matrix norm), then $U$ is said to be an $\epsilon$-approximated block encoding of $A$.
\end{definition}

The above definition has multiple natural corollaries. First, an arbitrary unitary $U$ block encodes itself. Suppose $A$ is block encoded by some matrix $U$. Next, then $A$ can be block encoded in a larger matrix by simply adding any ancilla (supposed to have dimension $m$), then note that $\Ibb_m \otimes U$ contains $A$ in the top-left corner, which is block encoding of $A$ again by definition. Further, it is almost trivial to block encode identity matrix of any dimension. For instance, we consider $\sigma_z \otimes \Ibb_m$ (for any $m$), which contains $\Ibb_m$ in the top-left corner. We further notice that from the above definition, the action of $U$ on some quantum state $\ket{\bf 0}\ket{\phi}$ is:
\begin{align}
    \label{eqn: action}
    U \ket{\bf 0}\ket{\phi} = \ket{\bf 0} A\ket{\phi} + \ket{\rm Garbage},
\end{align}
where $\ket{\rm Garbage }$ is a redundant state that is orthogonal to $\ket{\bf 0} A\ket{\phi}$.

\begin{lemma}[\cite{gilyen2019quantum} \revise{Block Encoding of a Density Matrix}]
\label{lemma: improveddme}
Let $\rho = \Tr_A \ket{\Phi}\bra{\Phi}$, where $\rho \in \mathbb{H}_B$, $\ket{\Phi} \in  \mathbb{H}_A \otimes \mathbb{H}_B$. Given unitary $U$ that generates $\ket{\Phi}$ from $\ket{\bf 0}_A \otimes \ket{\bf 0}_B$, then there exists a highly efficient procedure that constructs an exact unitary block encoding of $\rho$ using $U$ and $U^\dagger$ a single time, respectively.
\end{lemma}

The proof of the above lemma is given in \cite{gilyen2019quantum} (see their Lemma 45). \\



\begin{lemma}[Block Encoding of Product of Two Matrices]
\label{lemma: product}
    Given the unitary block encoding of two matrices $A_1$ and $A_2$, then there exists an efficient procedure that constructs a unitary block encoding of $A_1 A_2$ using each block encoding of $A_1,A_2$ one time. 
\end{lemma}

\begin{lemma}[\cite{camps2020approximate} \revise{Block Encoding of a Tensor Product}]
\label{lemma: tensorproduct}
    Given the unitary block encoding $\{U_i\}_{i=1}^m$ of multiple operators $\{M_i\}_{i=1}^m$ (assumed to be exact encoding), then, there is a procedure that produces the unitary block encoding operator of $\bigotimes_{i=1}^m M_i$, which requires \revise{parallel single uses} of 
    $\{U_i\}_{i=1}^m$ and $\mathcal{O}(1)$ SWAP gates. 
\end{lemma}
The above lemma is a result in \cite{camps2020approximate}. 
\begin{lemma}[\revise{\cite{gilyen2019quantum} Block Encoding of a  Matrix}]
\label{lemma: As}
    Given oracle access to $s$-sparse matrix $A$ of dimension $n\times n$, then an $\epsilon$-approximated unitary block encoding of $A/s$ can be prepared with gate/time complexity $\mathcal{O}\Big(\log n + \log^{2.5}(\frac{s^2}{\epsilon})\Big).$
\end{lemma}
This is presented in~\cite{gilyen2019quantum} (see their Lemma 48), and one can also find a review of the construction in~\cite{childs2017lecture}. We remark further that the scaling factor $s$ in the above lemma can be reduced by the preamplification method with further complexity $\mathcal{O}({s})$~\cite{gilyen2019quantum}.

\begin{lemma}[\cite{childs2012hamiltonian,gilyen2019quantum} Linear combination of block-encoded matrices]
    Given unitary block encoding of multiple operators $\{M_i\}_{i=1}^m$. Then, there is a procedure that produces a unitary block encoding operator of \,$\sum_{i=1}^m \pm M_i/m $ in complexity $\mathcal{O}(m)$, e.g., using block encoding of each operator $M_i$ a single time. 
    \label{lemma: sumencoding}
\end{lemma}

\begin{lemma}[Scaling Block encoding] 
\label{lemma: scale}
    Given a block encoding of some matrix $A$ (as in~\ref{def: blockencode}), then the block encoding of $A/p$ where $p > 1$ can be prepared with an extra $\mathcal{O}(1)$ cost.  
\end{lemma}
To show this, we note that the matrix representation of RY rotational gate is
\begin{align}
   R_Y(\theta) = \begin{pmatrix}
        \cos(\theta/2) & -\sin(\theta/2) \\
        \sin(\theta/2) & \cos(\theta/2) 
    \end{pmatrix}.
\end{align}
If we choose $\theta$ such that $\cos(\theta/2) = 1/p$, then Lemma~\ref{lemma: tensorproduct} allows us to construct block encoding of $R_Y(\theta) \otimes \mathbb{I}_{{\rm dim}(A)}$  (${\rm dim}(A)$ refers to dimension of matirx $A$), which contains the diagonal matrix of size ${\rm dim}(A) \times {\rm dim}(A)$ with entries $1/p$. Then Lemma~\ref{lemma: product} can construct block encoding of $(1/p) \ \mathbb{I}_{{\rm dim}(A)} \cdot A = A/p$.  \\

The following is called amplification technique:
\begin{lemma}[\cite{gilyen2019quantum} Theorem 30; \revise{\bf Amplification}]\label{lemma: amp_amp}
Let $U$, $\Pi$, $\widetilde{\Pi} \in {\rm End}(\mathcal{H}_U)$ be linear operators on $\mathcal{H}_U$ such that $U$ is a unitary, and $\Pi$, $\widetilde{\Pi}$ are orthogonal projectors. 
Let $\gamma>1$ and $\delta,\epsilon \in (0,\frac{1}{2})$. 
Suppose that $\widetilde{\Pi}U\Pi=W \Sigma V^\dagger=\sum_{i}\varsigma_i\ket{w_i}\bra{v_i}$ is a singular value decomposition. 
Then there is an $m= \mathcal{O} \Big(\frac{\gamma}{\delta}
\log \left(\frac{\gamma}{\epsilon} \right)\Big)$ and an efficiently computable $\Phi\in\mathbb{R}^m$ such that
\begin{equation}
\left(\bra{+}\otimes\widetilde{\Pi}_{\leq\frac{1-\delta}{\gamma}}\right)U_\Phi \left(\ket{+}\otimes\Pi_{\leq\frac{1-\delta}{\gamma}}\right)=\sum_{i\colon\varsigma_i\leq \frac{1-\delta}{\gamma} }\tilde{\varsigma}_i\ket{w_i}\bra{v_i} , \text{ where } \Big|\!\Big|\frac{\tilde{\varsigma}_i}{\gamma\varsigma_i}-1 \Big|\!\Big|\leq \epsilon.
\end{equation}
Moreover, $U_\Phi$ can be implemented using a single ancilla qubit with $m$ uses of $U$ and $U^\dagger$, $m$ uses of C$_\Pi$NOT and $m$ uses of C$_{\widetilde{\Pi}}$NOT gates and $m$ single qubit gates.
Here,
\begin{itemize}
\item C$_\Pi$NOT$:=X \otimes \Pi + I \otimes (I - \Pi)$ and a similar definition for C$_{\widetilde{\Pi}}$NOT; see Definition 2 in \cite{gilyen2019quantum},
\item $U_\Phi$: alternating phase modulation sequence; see Definition 15 in \cite{gilyen2019quantum},
\item $\Pi_{\leq \delta}$, $\widetilde{\Pi}_{\leq \delta}$: singular value threshold projectors; see Definition 24 in \cite{gilyen2019quantum}.
\end{itemize}
\end{lemma}
\end{document}